\documentclass[prc,superscriptaddress,showpacs,floatfix,twocolumn]{revtex4}
\usepackage[dvips]{graphicx}
\usepackage{epsfig}
\usepackage{pst-plot}
\usepackage{bm}
\usepackage{mathrsfs}
\usepackage{amsfonts}
\usepackage{amssymb}
\usepackage{lscape}
\usepackage{subfigure}

\usepackage{amsmath}
\begin{document}

\title{Gamow shell-model calculations of drip-line oxygen isotopes}
\author{K. Tsukiyama}
\affiliation{Department of Physics, University of Tokyo, 7-3-1 Hongo,
Bunkyo-ku, Tokyo, Japan}

\author{M.~Hjorth-Jensen}
\affiliation{Department of Physics and Center of Mathematics for Applications, 
University of Oslo, N-0316 Oslo, Norway}

\author{G. Hagen}
\affiliation{Physics Division, Oak Ridge National Laboratory, P.O. Box
2008, Oak Ridge, Tennessee 37831, USA}

\date{\today}
\begin{abstract}
We employ the Gamow shell model (GSM) to describe low-lying states of the 
oxygen isotopes $^{24}$O and $^{25}$O. The many-body Schr\"odinger 
equation is solved starting from a two-body Hamiltonian defined by 
a renormalized low-momentum nucleon-nucleon (NN) interaction, and a spherical
Berggren basis. The Berggren basis treats bound, resonant, and continuum
states on an equal footing, and is therefore an appropriate
representation of loosely bound and unbound nuclear states near
threshold. We show that such a basis is necessary in order to obtain
a detailed and correct description of the low-lying $1^+$ and $2^+$ 
excited states in $^{24}$O. On the other hand, we find that a correct description of
binding energy systematics of the ground states is driven by proper 
treatment and inclusion of many-body correlation effects. This is 
supported by the fact that we get $^{25}$O unstable with respect to 
$^{24}$O in both oscillator and Berggren representations starting from 
a $^{22}$O core. Furthermore, we show that the structure of these 
loosely bound or unbound isotopes are strongly influenced by the 
$^1S_0$ component of the NN interaction. This has 
important consequences for our understanding of nuclear stability.  
\end{abstract}
\maketitle

\emph{Introduction.} The study of nuclei far from stability is a  
leading direction in nuclear physics, experimentally and
theoretically. It represents a considerable intellectual challenge 
to our understanding of the stability of matter itself, with potential
implications for the synthesis of elements. An important aspect of 
this research direction is to understand how magic numbers and shells 
appear and evolve with increasing numbers of neutrons or protons. 
Except for a qualitative understanding of the neutron and proton 
dependence of the magic numbers in terms of mean-field models, we lack 
a quantitative theoretical understanding in terms of the basic 
constituents of the underlying nuclear many-body Hamiltonian. %After 
%more than five decades of nuclear physics research, we are not yet in 
%a position where we can quantitatively explain how shells evolve 
%from a fundamental point of view.  

For mesoscopic systems like nuclei, the interpretation of the standard
magic numbers $2$, $8$, $20$, $28$, $50$, $82$, and $126$ is linked to the  
interplay between the filling up of single-particle (s.p.) orbitals 
(s.p.~picture) and the underlying interactions. Magic numbers
lead to so-called shell gaps in the s.p.~spectra near the Fermi energy.  
%The nucleon that fills the last s.p. orbital below the Fermi level,  
%is normally well bound, while any additional nucleon is less bound, or even
%unstable with respect to particle emission. 
For various magic numbers, 
in particular for stable doubly magic nuclei, one can have a large
number of stable isotopes (increasing number of neutrons $N$ but 
fixed number of protons $Z$) or isotones  (increasing number of
protons $Z$ and fixed $N$). Eventually, as one adds more neutrons or 
protons, viz., moving away from the valley of nuclear stability
towards the drip lines, the outermost nucleons literally start to drip
off the nuclei, thereby defining the very limits of stable matter. 
%As one approaches these limits, new phenomena emerge such as
%halo densities, extreme matter clusterization and reorganization of shell structure,  
%as seen in the case of $N=20$ for nuclei around $^{32}$Mg, the so-called
%'island of inversion' region. Another example is the appearance of the 
%$N=14$ and $N=16$ shell gaps when one moves from the $N=20$ shell gap in $^{30}$Si to $^{24}$O. 

Nuclei near the drip line are open-quantum many-body systems
for which the coupling with the scattering continuum plays an
important role, and should therefore be explicitly taken
into account. Configuration interaction (shell-model) methods, such as
the Gamow shell model (GSM) (see Ref.~\cite{michel09} for a recent review of
the GSM) or the continuum shell model \cite{Volya06}, 
have been developed in order to properly include the coupling with the scattering continuum. 
The inclusion of continuum states complicates the
solution of the many-body problem considerably, as the number of many-body
basis states will explode. This has therefore motivated different
approaches to the description of loosely bound and unbound nuclear states,  
such as the Density-Matrix-Renormalization-Group (DMRG) \cite{rotureau06} and 
Coupled-Cluster approaches \cite{hagen07}.
% These methods scale much more gently with the
%system size, and allow for an explicit inclusion of the scattering continuum.    
Furthermore, to complicate the solution strategies of the nuclear
many-body problem is the need to include three-nucleon forces (3NFs). 
Recent results for light nuclei using \emph{ab-initio} methods such as the Green's
function Monte-Carlo (GFMC) \cite{pieper02} and the no-core shell
model (NCSM) \cite{navratil02}, demonstrate that 3NFs are needed in 
order to give the correct binding energies and spectroscopy when 
comparing to experiment. It is still an open research problem what 
the role of 3NFs is in medium mass and neutron-rich nuclei close 
to the drip line. This applies also to our understanding of
 how different parts of 
the underlying nucleon-nucleon (NN) interaction, such as the spin-orbit 
force \cite{schiffer2004,warner2004} and the tensor force
\cite{taka2005,taka2001}, affect the structure of nuclei close to the drip line.

In this work we study the ground- and low-lying states of $^{24}$O
and $^{25}$O within the GSM framework. We derive for the first time 
a realistic effective shell-model interaction that includes 
the effect of the scattering continuum. The effective interaction is derived using 
many-body perturbation theory (MBPT) starting with a Hamiltonian which 
reproduces NN scattering data. The choice of the oxygen isotopes is motivated by
several reasons. First, the oxygen isotopes are the heaviest isotopes for 
which the drip line is well established. There are large experimental 
campaigns worldwide \cite{elekes2007,schiller2006,hoffman2008} which
aim at uncovering the properties of the oxygen isotopes, both at or
close to the drip line and beyond. Two out of four stable even-even
isotopes exhibit a doubly magic nature, namely $^{22}$O ($Z$=$8$,$N$=$14$) 
\cite{thirolf2000} and $^{24}$O ($Z$=$8$, $N$=$16$) \cite{stanoiu2004,hoffman2009}.   
%The structure of these two doubly magic nuclei is assumed to be governed 
%by the evolution of the $1s_{1/2}$ and $0d_{5/2}$  one-quasiparticle states.  
The isotopes $^{25-28}$O %, $^{26}$O, $^{27}$O and $^{28}$O 
are all
believed to be unstable towards neutron emission, even though $^{28}$O
is a doubly magic nucleus within the standard shell-model picture. 
%It 
%is remarkable that one more proton can bind six more neutrons in the fluorine
%isotope chain, shifting the dripline all the way out to $^{31}$F \cite{Sakurai99}.   
% \include refs to  many expt groups
Secondly, many-body descriptions starting with effective two-body
Hamiltonians defined for a model space comprising the $1s_{1/2}$,
$0d_{5/2}$, and $0d_{3/2}$ s.p.~orbitals, have always defied a proper
reproduction of the experimental data \cite{brown2005,brown2007,mhj95}. 
To remedy this, the two-body shell-model interaction is usually fitted 
(fully or partially) \cite{brown2005,brown2007,coraggio2007,Volya06} to
reproduce experimental data for nuclei in the range $16 <  A < 40$. 
Without fitting the two-body Hamiltonians, which are typically derived 
using MBPT, shell-model calculations of the 
oxygen isotopes fare reasonably up to $^{20}$O, but produce too
compressed spectra in $^{22}$O and lead to neutron separation energies 
that are always positive up to $^{28}$O. All isotopes $^{17-28}$O
that can be reached via this model space result as well bound, in
clear contradiction with experiment. These results pertain to all 
available models of the NN interaction \cite{brown2005,mhj95}
within the above model space. The ground states and excited states of
the odd-even isotopes are also poorly reproduced (see for example 
Refs.~\cite{brown2005,brown2007}).

%It is clear that this poor description of the heavier oxygen
%isotopes is mainly due to a severely limited model space and
%improper treatment of many-body correlations. Effective three- and many-body 
%forces will therefore play a significant role in a realistic description of the oxygen isotopes. 
%In shell-model calculations starting from a $^{16}$O core, these
%many-body effects are further enhanced as the number of neutrons
%increases. 
Recent \emph{ab-initio} coupled-cluster calculations of the oxygen
isotopes \cite{hagen4} showed that chiral NN interactions 
can produce a rather flat binding energy curve of the oxygen isotopes
ranging from $^{24}$O to $^{28}$O. 
It was illustrated that the existence of $^{28}$O cannot be ruled out
from \emph{ab-initio} theory starting from modern chiral
interactions. It was concluded that 3NFs will eventually decide the
matter. Recent shell-model calculations of the oxygen
isotopes within the \emph{s-d} shell showed that inclusion of
effective 3NFs give added repulsion in the heavier oxygen isotopes and
resulted in an unstable $^{28}$O \cite{taka09}.  

Obviously, the understanding of the evolution of binding energies 
and the location of the drip line in the oxygen isotopes 
are still unsolved theoretical problems. As shown in
Refs.~\cite{hagen4,taka09}, proper treatment of many-body effects are
crucial in the oxygen isotopes. However, it is still an open issue 
what the role of the continuum is in the evolution of shell structure 
and in a correct description of binding energy systematics in neutron-rich 
oxygen isotopes. 
%The abovementioned shell model calculations of the oxygen
%isotopes assume that the$0d_{3/2}$ s.p. orbit $^{17}$O is bound, 
%whereas experimentally there is a well-known resonance in $^{17}$O
%with quantum numbers $3/2^+$. Since one normally employs harmonic
%oscillator wave functions to represent the s.p. wave
%functions, one introduces the wrong asymptotics in constructing an 
%effective shell model interaction. In Ref. \cite{volya06} the
%continuum shell model approach was applied to the oxygen isotopes,
%employing a phenomenological interaction a qualitatively correct
%spectrum was obtained. 
%Ideally one should construct the effective
%hell model interaction for the oxygen isotopes starting from a basis 
%which includes the discretized continuum and $0d_{3/2}$ as a resonant
%state. 
As found in Ref.~\cite{quaglioni08}, a proper treatment of the
continuum was necessary in order to explain the parity inversion of
the $^{11}$Be ground state. It is therefore important to investigate 
what the role of continuum is on ground and excited states in
neutron-rich oxygen isotopes. It is the aim of this article to throw light on, 
and disentangle the different effects coming from many-body correlations and
the proximity of the continuum on the evolution of binding energies
and shell structure in neutron-rich oxygen isotopes. Further, we
will investigate how different components of the NN
interaction influence the low-lying resonance states in drip-line oxygen isotopes.

\emph{Method, interaction, and model space.} In this work we use the
GSM to describe the low-lying states in $^{24}$O and the ground state 
of $^{25}$O. We construct the effective shell-model interaction
starting from $^{22}$O as a closed-shell core and use MBPT through second order
\cite{mhj95}. The isotope $^{22}$O has been found to be a closed-shell nucleus
with a considerable shell gap, making it suitable as a starting point for
shell-model calculations. With $^{22}$O as a closed-shell
core we have at most three valence particles, making
effects coming from a proper treatment of the continuum more
transparent. So far, the effective interaction has typically been 
constructed using oscillator states as intermediate particle states. 
In this work, we investigate for the first time what the role of
the continuum is in constructing a realistic effective shell-model 
interaction and in describing spectra of neutron-rich 
oxygen isotopes.

The major advantage of the GSM is that it unifies
structure and reaction properties of nuclei, that is, bound and
scattering degrees of freedom are treated equally. Nuclei close to
the scattering threshold have low-lying states that are in the
continuum, and the GSM provides a simple framework for
calculating the energy and lifetimes of these states. The starting
point of the GSM is the Berggren completeness relation
\cite{Berggren1968}. %,Berggren1971,Berggren1978,Berggren1996,lind}.
This s.p.~basis treats bound, resonant, and scattering states on an equal
footing, and it is the basic idea of the GSM to expand the
many-body wave function in Slater determinants built from this basis.
The representation of the Hamiltonian will no longer be Hermitian but
rather complex symmetric, and a diagonalization will yield bound
state spectra in addition to resonant state spectra. Hence, the energy
and lifetime of the many-body resonant state comes out directly from the
approach.   

In constructing the s.p.~Berggren basis, we start from the 
one-body Hamiltonian for a spherical Woods-Saxon (WS) potential (see
for example Ref.~\cite{witek2}). The Berggren basis is obtained by diagonalizing
the one-body Hamiltonian in a plane wave basis defined on a deformed
contour $L^+$ in the complex momentum plane (see Ref.~\cite{hagen3} for
details). This method is also known as the Contour-Deformation-Method.
There is freedom in choosing the contour $L^+$ provided the
Schr\"odinger equation can be analytically continued in the
complex momentum plane. Fully converged results should be independent 
of the choice of the contour, and in the following we used two different
contours, $L_1^+$ and $L_2^+$, as given in Ref.~\cite{hagen3}, to check that
our results are converged. The parameters of the WS potential are
given by $V_0=55$ MeV, $V_{so}=29$ MeV, $R=3.1$ fm, and $d=0.45$ fm, where the shorthand $V_{so}=29$
stands for the spin-orbit term. These parameters are chosen such that
the energies of the s.p.~orbitals $\nu 0d_{5/2}$, $\nu 1s_{1/2}$, and $\nu
0d_{3/2}$ are close to the experimental values obtained by taking the
binding energy differences of $^{21}$O-$^{22}$O and $^{23}$O-$^{22}$O. 
We obtain, in units of MeV, $\epsilon_{\nu 0d5/2}$=$-6.53$($-6.85$), $\epsilon_{\nu 1s1/2}$
=$-2.73$($-2.74$) and $\epsilon_{\nu d3/2}$=$1.43-0.01i$ ($1.26-0.2i$),
where numbers in parenthesis are experimental values. We refer
hereafter to this basis as a WS Gamow basis. In our shell-model study
of the oxygen isotopes $^{24}$O and $^{25}$O we will use the WS Gamow basis
for the neutron partial waves $s_{1/2}$, $d_{3/2}$, and $d_{5/2}$, 
while for protons and all other partial waves, we use the oscillator 
representation.\\
\hspace*{11pt}In constructing the effective shell-model interaction for neutron-rich
oxygen isotopes, we start from the intrinsic nuclear Hamiltonian $ 
H = t - t_{\mathrm{CoM}} + V $, where $t$ denotes the kinetic energy,
and $t_{\rm CoM}$ is the kinetic energy of the center of mass. 
For the interaction $V$, we employ the NN interaction by Entem and
Machleidt~\cite{machleidt02,Entem2003} derived from chiral effective
field theory (EFT) at next-to-next-to-next-to-leading order (N$^3$LO)
and with a 500 MeV cutoff. This interaction still has
considerable high-momentum components, making MBPT
non-convergent. In order to construct the effective shell-model
interaction $V_{\rm eff}$ using MBPT, we use a
low-momentum interaction $V$=$V_{{\rm low-}k}$, obtained by integrating out those momentum 
components above a certain cutoff $\Lambda$ \cite{bogner2003}. 
Thereafter we generate $V$ in the WS Gamow basis using the procedure outlined in
Ref.~\cite{hagen3}. It is well known that this renormalization 
induces forces of higher rank. In a completely renormalized theory, 
all observables are independent on the cutoff $\Lambda$, and therefore 
the dependence of our results on the cutoff $\Lambda $ tells us about 
the missing many-body physics. In this work we start from a $^{22}$O
core and limit ourselves to two-body effective
shell-model interactions only. Since we have at most three
neutrons outside the closed-shell core, it is reasonable to expect 
that the effect coming from effective three-body forces is small.

\begin{figure}[htbp]
 \includegraphics[width=0.3\textwidth]{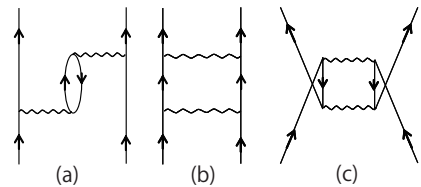}
\caption{The second-order contributions to $V_{\rm eff}$.}
\label{fig:q2}
\end{figure}

In order to proceed with the construction of the effective shell-model
interaction, we need to define a model space for the valence
neutrons. Since we will perform our GSM calculations starting with 
$^{22}$O as a closed shell core, i.e., the $0d_{5/2}$ shell is filled,
our many-body states are constructed by letting all valence
neutrons act in the model space defined by the the 
s.p.~orbitals $1s_{1/2}$ and $id_{3/2}$, where $i=0,1,\dots$ denote the
$i$-th discretized continuum state for the $d_{3/2}$ partial wave. 
The effect of the continuum from the $s_{1/2}$ and $d_{5/2}$ partial
waves are taken into account through the construction of the effective interaction. 
The effective shell-model interaction $V_{\rm eff}$ is then constructed for this
model space using MBPT \cite{mhj95},
\begin{align*}
 H_{eff}&=\hat{Q}-\hat{Q}'\int \hat{Q}+\hat{Q}'\int\hat{Q}\int\hat{Q}-\cdots.
\end{align*}
Here $\hat{Q}(\omega)$ is given by diagrams which are valence-linked and
irreducible. The argument $\omega$ is the starting energy, which is the
sum of the unperturbed s.p.~energies. The object
$\hat{Q}(\omega)$ is defined as 
\begin{align}
 \hat{Q}(\omega)&=PH_1P+PH_1Q\frac{1}{\omega -QHQ}QH_1P. \nonumber
\end{align}
In this work we take into account diagrams up through second order in 
perturbation theory in order to construct $\hat{Q}(\omega)$. 
%The
%effective shell model interaction to second order in perturbation theory can
%then be written in diagrammatic form as, 
In Fig.~\ref{fig:q2} the diagrams that contribute up to second order
in the effective interaction are shown.  
The intermediate s.p.~states in each diagram in
can be either resonant, non-resonant continuum,
or bound states. In our calculations we neglect the contributions 
from folded diagrams; see Ref.~\cite{mhj95} for further details. 

\emph{Results.} We now turn to the GSM calculations of
low-lying states in $^{24}$O and $^{25}$O. For the case of $^{24}$O we
have two neutrons outside the $^{22}$O core, and we express the ground
state of $^{24}$O by filling the $1s_{1/2}$ orbital. This orbital is
well bound, and with a neutron separation energy of $S_n$=2.7 MeV for
$^{23}$O \cite{masstable}. We therefore expect mixing of higher-lying 
states to be small in the ground state of $^{24}$O, and we take
this effect into account through the construction of the effective 
interaction using many-body perturbation theory. For the excited states $J=1$ 
and $J=2$ of $^{24}$O, we expand the wave function in a set of 
basis states $|iJ^+\rangle =|1s_{1/2}\otimes id_{3/2};J^+
\rangle$. The orbital $id_{3/2}$ denote the $i$-th discretized
s.p.~state of the WS Gamow basis, and $i$=$1,\cdots n_{\rm gamow}$, where
$n_{\rm gamow}$ is typically between $20$ and $35$. In our calculations
we used $n_{\rm gamow}$=$30$, and we checked that this was sufficient
to reach convergence. 

\begin{figure}[htbp]
 \includegraphics[width=.4\textwidth]{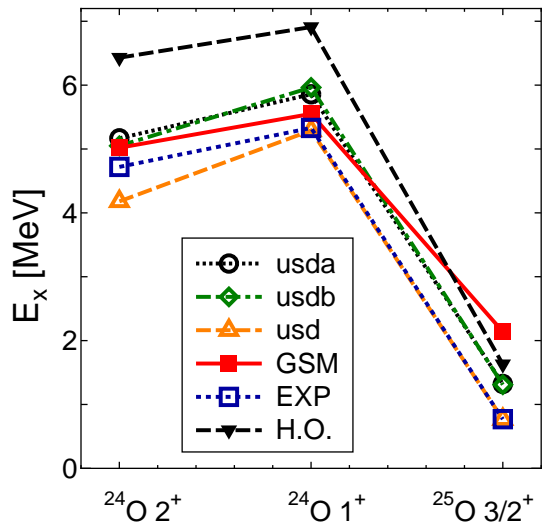}
\caption{The excitation energies of the $J^{\pi}=1^+$ and $J^{\pi}=2^+$ states in $^{24}$O and the
 ground-state energy of $^{25}$O measured from the ground states in
 $^{24}$O. }
\label{fig:low-lying}
\end{figure}

In Fig.~\ref{fig:low-lying} the excitation energies of the low-lying 
states in $^{24}$O and the ground state in $^{25}$O with respect to
the ground state of $^{24}$O are shown. In these calculations we employed a
cutoff $\Lambda = 2.1 \mathrm{fm}^{-1}$ for the $V_{{\rm low-}k}$
interaction. We have varied the cutoff $\Lambda$ around $2.1$ fm$^{-1}$ and
found that our results are nearly independent of the cutoff $\Lambda$, 
meaning that the effect of neglected three- and many-body forces are small
and will not change our conclusions. Our GSM calculations give energies for the 
low-lying states that compare well with the experimental values. 
The splitting between the $1^+$ and the $2^+$ states of $^{24}$O 
compares well with experiment. The theoretical spacing obtained with
our GSM approach is 540 keV, while the corresponding
experimental value is 600 keV. Our calculated widths for the
$1^+$ and $2^+$ resonant states are $0.44$ and $0.26$ MeV,
respectively. In Ref.~\cite{hoffman2008}, Hoffman reported 
that the widths are $0.03^{+0.12}_{-0.03}$ MeV for
the $1^+$ state and $0.05^{+0.21}_{-0.05}$MeV for the $2^+$ state.
Our calculations overestimate the width of the $1^+$ state as compared
to experiment, on the other hand our calculated width of the  $2^+$
state is within the experimental uncertainties. 
%Koshiroh: Do we have experimental values for the lifetimes or widths
%of the 1+ and the 2+ states in O24 ? 
%
% Gaute: the widths for 1+ and 2+ states are not exactly determined.
% C. Hoffmann reported in PLB 672, 17 (2008) that gamma are
% gamma<0.26 MeV for 1+ and gamma<0.15 MeV for 2+. These are based on
%R-matrix theory and have uncertainty.  
As a comparison, we show results
obtained with other effective interactions. The new universal
{\it sd} Hamiltonians (USDA and USDB), obtained by revising the USD 
interaction in order to be more suitable for exotic nuclei
\cite{brown2007}, results in a good agreement with experimental data. 
However, one should note that some effects coming from the continuum and 
3NFs are implicitly included in the new USD two-body matrix elements due to 
the fit made to experimental data. Although the USD interaction reproduces 
nicely two of the three experimental data points, the $2^+$ state in $^{24}$O 
comes out as a bound excited state, a fact which disagrees  with
recent experiments. We also show results obtained with a pure
harmonic oscillator basis. In this case, the discretized continuum is
not included in the model space, and we construct the effective
interaction using perturbation theory through second order (see
Fig.~\ref{fig:q2}), and a model space consisting of neutrons in the 
orbitals $(1s_{1/2},0d_{3/2})$. Using a model space consisting of only bound
oscillator orbitals puts the $1^+$ and $2^+$ states in $^{24}$O too
high in energy when comparing with experiment. 
%Koshiroh: In order to compare more directly with our GSM calculations 
%can you increase the number of d3/2 oscillator orbitals in the
%s.p. model space ?
With the inclusion of the continuum in the model space, the $1^+$ and $2^+$ states get lower in
energy and also closer to experiment. This clearly shows an effect coming
from a proper treatment of the continuum, and should therefore be 
included in a realistic description of low-lying states in $^{24}$O. 
Further, the $2^+$ state in $^{24}$O is a resonant state and cannot 
be properly described in an oscillator basis. 

Turning to the case of the ground state of $^{25}$O, our GSM calculations 
correctly predict that it is particle unstable, but puts it at a
higher energy ($E$=$2.1$, $\Gamma$=$0.5$ MeV) as compared to recent
experimental data ($E$=$0.77$, $\Gamma$=$0.17$ MeV). To get a better description of this
state, configurations like $\nu (1s_{1/2})(id_{3/2})^2$ should be
included since the GSM is currently solved in a filling configuration.
% Koshiroh: I though that $\nu (1s_{1/2})(id_{3/2})^2$ configurations
% were already included by the definition of our model space.  
% I cant see where you say that these configurations are excluded. And
% why is that ?
Again we compare our GSM result for $^{25}$O with a shell-model
calculation where only the bound oscillator orbitals
$(1s_{1/2},0d_{3/2})$ define the model space. An important point to
note is that $^{25}$O comes out higher in energy as compared to
$^{24}$O, reflecting that it is unstable towards neutron emission. 
Even though the oscillator basis cannot provide us with a width or a lifetime
for the ground state of $^{25}$O, it predicts $^{25}$O as particle
unstable and illustrates that the continuum coupling is not a dominant
mechanism in explaining the position of the drip line of the oxygen isotopes.  
Thus, irrespective of whether we use a complex Gamow s.p.~basis that defines a model space consisting of the
$(1s_{1/2},id_{3/2})$ states (with $i=0,\dots, n_{\rm gamow}$), or a
harmonic oscillator basis with the $(1s_{1/2},0d_{3/2})$ states only, $^{25}$O
is unstable against particle emission as long as we use $^{22}$O as a
closed-shell core. This has important consequences for our theoretical 
understanding of the stability of matter. As pointed out in a recent 
calculation by Hagen {\em et al} \cite{hagen4}, many-body forces such 
as three-body interactions are crucial in explaining the correct
position of the drip lines. The fact that a calculation of $^{25}$O 
with $^{22}$O as a closed-shell core gives the correct qualitative 
description of the ground state of $^{25}$O, hints at the fact that
with three valence neutrons only, missing many-body physics does not 
play a central role. This should be contrasted to realistic calculations that 
utilize $^{16}$O as a closed core. With nine valence neutrons, $^{25}$O
is strongly overbound using the standard model space comprising the
oscillator orbitals $(1s_{1/2},0d_{5/2},0d_{3/2})$ and a two-body effective
interaction derived using MBPT. However, 
the location of the excited states in $^{24}$O and the width of the
ground state of $^{25}$O depend crucially on the choice of basis, and
we have shown that a GSM basis is needed in order to
explain these states. 
  
\begin{figure*}[htbp]
 \subfigure[The ground state and the $1^+$ in $^{24}$O]{\includegraphics[width=.4\textwidth]{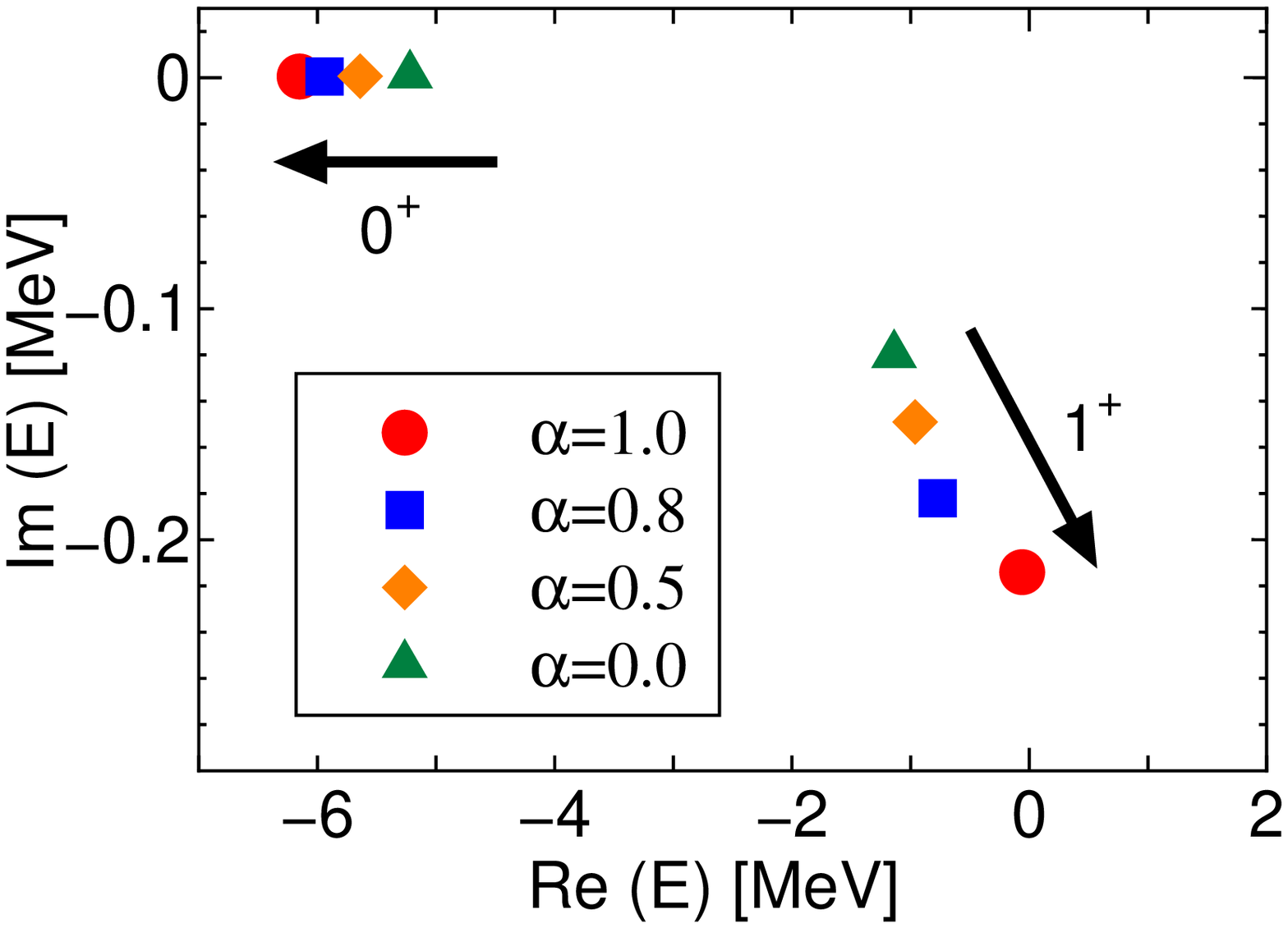}}
 \subfigure[The ground state and the $2^+$ in $^{24}$O]{\includegraphics[width=.4\textwidth]{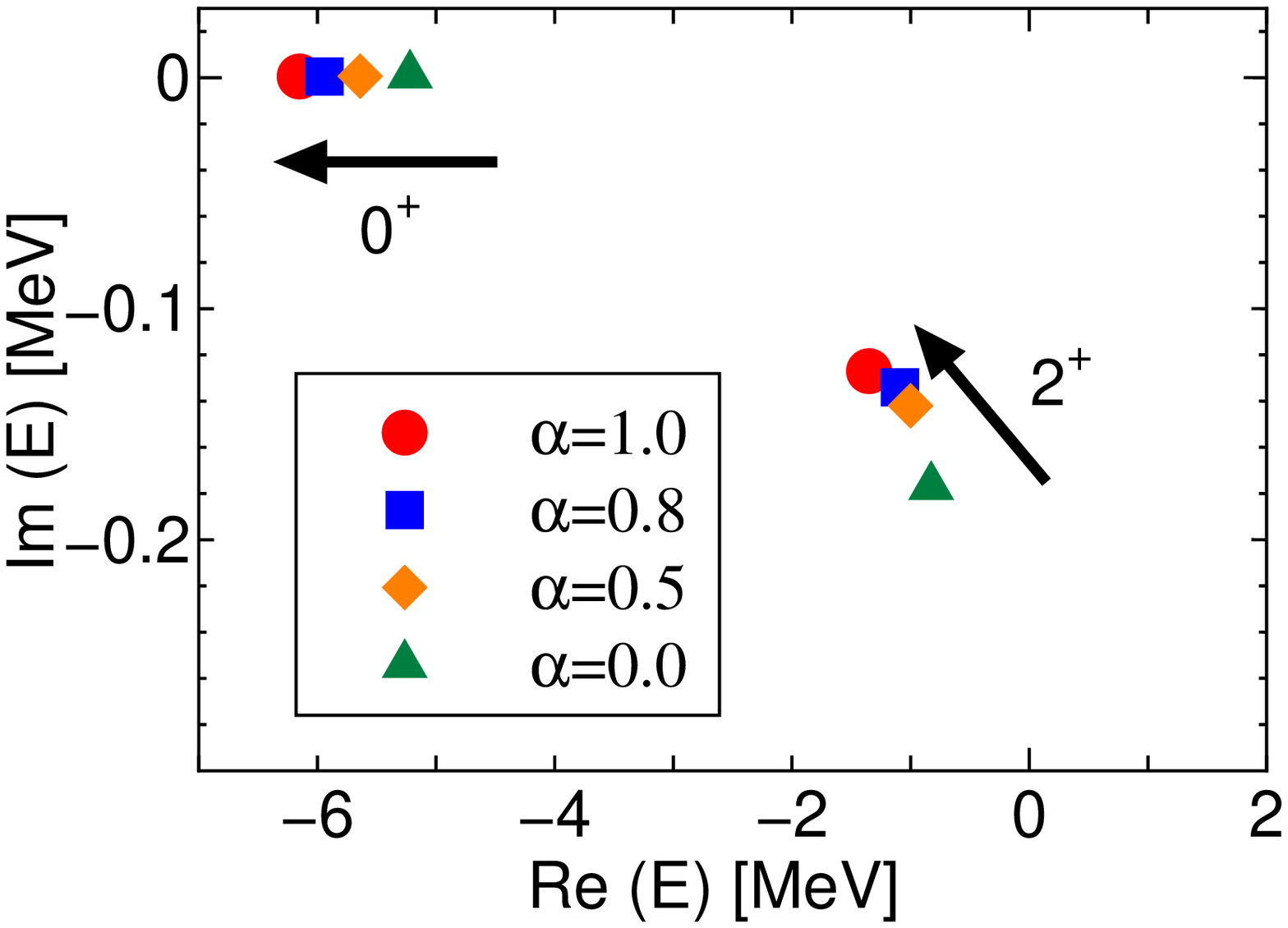}}
\caption{The energies of low-lying resonance states in $^{24}$O, as a
 function of the strength $\alpha$ of the $^1S_0$ partial wave component of the nucleon-nucleon  interaction. See text for
 further details.}
\label{fig:paring}
\end{figure*}

We conclude this work with an investigation of which parts of the 
NN interaction are crucial for an understanding
of the above excited states in $^{24}$O. To achieve this, we single   
out the $^1S_0$ partial wave component of the NN
interaction and vary its strength. This partial wave is particularly 
important to our understanding of pairing correlations in nuclei.
In order to investigate the importance of this partial wave on the 
low-lying resonance states in $^{24}$O, we vary the strength as 
$ \tilde{V}(\alpha)$=$\alpha\times\langle 
S$=$0,L$=$0;J$=$0|V|S$=$0,L$=$0;J$=$0\rangle, \alpha\in [0,1]$ in a bare
NN potential. 
Clearly, $\alpha$=$0$ corresponds to an NN interaction
with no contribution from this partial wave while $\alpha$=$1$ corresponds to the original
interaction. The low-lying resonance states are calculated for each
$\tilde{V}(\alpha)$, and the results are shown in
Fig.~\ref{fig:paring}. One can clearly see how the $^{1}S_0$ partial wave in
the NN interaction influences the many-body resonances in 
$^{24}$O. The ground state gains additional binding when $\alpha$ 
increases since the ground state can be well described by the $\nu(1s_{1/2})^2$
configuration. At $\alpha$=$0$ the position of $1^+$ and $2^+$ are 
interchanged  with respect to experiment, 
and they are degenerate if one analyzes
the real part of the resonance energies. When $\alpha$ increases to
the full interaction strength, the $1^+$-$2^+$ splitting increases to
$530$ keV, a result close to experiment. The $^{1}S_0$ partial wave, a
central contribution to pairing in nuclei, is thus crucial  in order to explain 
correctly the low-lying resonant states in $^{24}$O.

\emph{Conclusions.} We have described the low-lying states in the
oxygen isotopes $^{24}$O and $^{25}$O within the framework of the
GSM and effective interactions derived through many-body
perturbation theory. We have shown that there is a fine interplay
between many-body effects and the proximity of the continuum. 
We demonstrated that the inclusion of a 
discretized continuum and resonant states
are crucial in order to describe the excited states of $^{24}$O
and the ground state of $^{25}$O. A Gamow basis is needed in order
to reproduce both the correct excitation energies and widths of these
states. In order to predict the location of the drip line, we found
that a proper inclusion of many-body forces and correlation effects are
the deciding factors. Furthermore, the $^1S_0$ partial wave component 
of the NN interaction, an important contribution to
strong pairing correlations in nuclei, plays an important role in
explaining the  correct value of these states. 
We conclude that the location
of the drip line and excited states in the oxygen isotopes can only be 
determined by proper inclusion of both many-body correlations and the 
coupling with continuum degrees of freedom.

%
% Acknowledgment
%

We thank Prof. T. Otsuka for valuable comments.
This work has been supported in part by the JSPS core-to-core program, 
EFES and by a grant-in-aid for Scientific Research (A) 20244022, and by the
Research Council of Norway (Supercomputing grant NN2977K).  
Oak Ridge National Laboratory is managed by UT-Batelle, LLC, for the
U.S.~Department of Energy under Contract No. DE-AC05-00OR22725.
KT thanks JSPS for a fellowship.


\begin{thebibliography}{100}

\bibitem{michel09}
N. Michel {\em et al}, % W. Nazarewicz, M. Ploszajczak, T. Vertse, 
J.~Phys.~G {\bf 36}, 013101 (2009).

\bibitem{Volya06}
A. Volya and V. Zelevinsky, \prc {\bf 74}, 064314 (2006).


\bibitem{rotureau06}
J. Rotureau {\em et al}, %N. Michel, W. Nazarewicz, M. Ploszajczak, J. Dukelsky, 
Phys.~Rev.~Lett.~{\bf 97}, 110603 (2006).

\bibitem{hagen07}
G. Hagen {\em et al}, %D. J. Dean, M. Hjorth-Jensen and T. Papenbrock, 
Phys.~Lett.~B, {\bf 656} (4), 169-173, (2007).

\bibitem{pieper02}
R.~B.~Wiringa, Steven~C.~Pieper, 
Phys.~Rev.~Lett. {\bf 89}, 182501 (2002).  

\bibitem{navratil02}
P. Navratil and W. E. Ormand, 
Phys. Rev. Lett. {\bf 88}, 152502 (2002).

\bibitem{schiffer2004}
J.~P.~Schiffer Phys.~Rev.~Lett.~{\bf 92}, 162501 (2004).

\bibitem{warner2004}
D.~Warner, Nature {\bf 430}, 517 (2004).

\bibitem{taka2005}
T.~Otsuka {\em et al}, %T.~Suzuki, R.~Fujimoto, H.~Grawe, and Y.~Akaishi, 
Phys.~Rev.~Lett.~{\bf  95}, 232502 (2005).

\bibitem{taka2001}
T.~Otsuka {\em et al}, %R.~Fujimoto, Y.~Utsuno, B.~A.~Brown, M.~Honma,
%  and T.~Mizusaki, 
Phys.~Rev.~Lett.~{\bf 87}, 082502 (2001).

\bibitem{elekes2007}
Z.~Elekes {\em et al}, Phys.~Rev.~Lett.~{\bf 98}, 102502 (2007).

\bibitem{schiller2006}
A.~Schiller {\em et al}, Phys.~Rev.~Lett.~{\bf 98}, 102502 (2007).

\bibitem{hoffman2008}
C.~R.~Hoffman {\em et al}, Phys.~Rev.~Lett.~{\bf 100}, 152502 (2008).


\bibitem{thirolf2000}
P.~G.~Thirolf {\em et al}, Phys.~Lett.~{\bf B485}, 16 (2000).

\bibitem{stanoiu2004}
M.~Stanoiu {\em et al}, Phys.~Rev.~C {\bf 69}, 034312 (2004).

\bibitem{hoffman2009}
C.~R.~Hoffman {\em et al}, Phys.~Lett.~{\bf B672}, 17 (2009).

\bibitem{Sakurai99}
H. Sakurai {\it et al.}, Phys. Lett. B {\bf 448}, 180 (1999).

\bibitem{brown2005}
B.~A.~Brown and W.~A.~Richter, Phys.~Rev.~C  {\bf 72}, 057301 (2005).

\bibitem{brown2007}
B.~A.~Brown and W.~A.~Richter, Phys.~Rev.~C  {\bf 74}, 034315 (2006).

\bibitem{mhj95}
M.~Hjorth-Jensen, T.~T.~S. Kuo, and E.~Osnes,  Phys.~Rep.~{\bf 261}, 125 (1995).

\bibitem{coraggio2007}
L.~Coraggio {\em et al}, %A.~Covello, A.~Gargano, N.~Itaco, D.~R. Entem, T.~T.~S.~Kuo, and
%  R.~Machleidt.
Phys.~Rev.~C  {\bf 75}, 024311 (2007).

\bibitem{hagen4} G.~Hagen {\em et al},
% T.~Papenbrock, D.~J.~Dean, M.~Hjorth-Jensen, B.~Velamur Asokan, 
Phys.~Rev.~{\bf 80}, 021306(R) (2009).

\bibitem{taka09}
Takaharu Otsuka {\em et al}, arXiv:0908.2607 (2009). 


\bibitem{quaglioni08} 
Sofia Quaglioni and Petr Navratil, 
Phys.~Rev.~Lett.~{\bf 101}, 092501 (2008).

% \bibitem{michel1} N.~Michel, W.~Nazarewicz and M.~P{\l}oszajczak,
% Phys.~Rev.~C {\bf 70},  064313 (2004)

% \bibitem{michel2} N.~Michel, W.~Nazarewicz, M.~P{\l}oszajczak and J.~Rotureau,
% Rev.~Mex.~Fis.~{\bf 50}  Suppl.~2, 74 (2004).


% \bibitem{witek1}N.~Michel, W.~Nazarewicz, M.~P{\l}oszajczak, and
% 	K.~Bennaceur, Phys.~Rev.~Lett.~{\bf 89}  042502 (2002).

 \bibitem{witek2}N.~Michel {\em et al}, 
%W.~Nazarewicz, M.~P{\l}oszajczak, and
% 	J.~Oko{\l}owicz, 
Phys.~Rev.~C {\bf 67},  054311 (2003).

% \bibitem{roberto}R.~Id Betan,  R.~J.~Liotta, N.~Sandulescu, and
% 	T.~Vertse, Phys.~Rev.~Lett.~{\bf 89}  042501 (2002).
% \bibitem{betan}R.~IdBetan  R.~J.~Liotta, N.~Sandulescu, and T.~Vertse, 
% 	Phys.~Rev.~C {\bf 67},  014322 (2003).

% \bibitem{betan2}R.~Id Betan  R.~J.~Liotta, N.~Sandulescu, and T.~Vertse, 
% 	Phys.~Lett.~{\bf B584},  48 (2004)

% \bibitem{hagen1}G.~Hagen, J.~S.~Vaagen, and M.~Hjorth-Jensen,
% J.~Phys.~A,  Math.~Gen.~{\bf 37},  8991 (2004).

% \bibitem{hagen2}G.~Hagen, M.~Hjorth-Jensen, and J.~S.~Vaagen,
% Phys.~Rev.~C {\bf 71},  044314 (2005).
\bibitem{bogner2003}
S. K. Bogner, T. T. S. Kuo and A. Schwenk,
Phys. Rep.~{\bf 386}, 1 (2003)

\bibitem{hagen3}G.~Hagen, M.~Hjorth-Jensen, and N.~Michel,
Phys.~Rev.~C {\bf 73},  044314 (2006).

\bibitem{Berggren1968}
T.~Berggren, Nucl.~Phys.~{\bf A109}, 265 (1968).

% \bibitem{Berggren1971}
% T.~Berggren, Nucl.~Phys.~{\bf A169}, 353 (1971).

% \bibitem{Berggren1978}
% T.~Berggren, Phys.~Lett.~{\bf B73}, 389 (1978).

% \bibitem{Berggren1996}
% T.~Berggren, Phys.~Lett.~{\bf B373}, 1 (1996).

% \bibitem{lind}P.~Lind, Phys.~Rev.~C {\bf 47},  1903 (1993).

\bibitem{machleidt02} D.~R.~Entem and R.~Machleidt, Phys.~Lett.~{\bf B524}, 93 (2002).

\bibitem{Entem2003}
D.~R.~Entem and R.~Machleidt,  Phys.~Rev.~C  {\bf 68}, 41001 (2003).

%\bibitem{bogner2007}
%S.~K.~Bogner, R.~J.~Furnstahl, and R.~J.~Perry, Phys.~Rev.~C {\bf 75}, 061001 (2007).

\bibitem{masstable} G.~Audi, A.~H.~Wapstra and C.~Thibault, Nucl.~Phys.~{\bf A729},  337 (2003).

\end{thebibliography}
\end{document}